# Air Cherenkov Methods in Cosmic Rays: A Review and Some History[1]

A.S. Lidvansky

Institute for Nuclear Research, Russian Academy of Sciences, 60th October Anniversary pr. 7a, Moscow, 119312 Russia

e-mail: lidvansk@sci.lebedev.ru
fax: +7(095)1358560

## Introduction

> "Always a modest individual, he was extremely scrupulous not to pretend to be involved in the developing applications just because of his contribution to the effect discovery. He even may have avoided using the Cherenkov technique in his own experiments."
>
> A.E. Chudakov, Pavel Alekseyevich Cherenkov (Obituary), Physics Today, December 1992

Indeed, the 'glorious development of the Cherenkov technique in experimental physics' (Chudakov, 1992) proceeded almost without the participation of the discoverer who gave his name to this radiation. In the field of cosmic ray studies the application of Vavilov-Cherenkov radiation can be subdivided into three large branches. One is the application of detectors with Cherenkov radiators in the instruments onboard spacecraft (either orbiting around the Earth or sent to deep space) and balloons. The second branch is related to large-volume water (or ice) Cherenkov detectors measuring the fluxes of neutrinos and muons under a thick overburden of rock, water, or ice (two types of these detectors make use of specially constructed water capacities and natural media). The third branch (historically it began to be developed first) uses the atmosphere as a radiator of Cherenkov light. Accordingly, the experimental methods here deal with collection of the light generated in air by natural fluxes of very high-energy cosmic ray particles.

It is worth noting that in the first case the use of the Cherenkov technique has much in common with the application of similar detectors in different areas of physics (for example, at accelerators). On the contrary, the two latter branches are characteristic of their own very specific instrumentation and methods. This paper is devoted exclusively to the third issue. By definition, observations here are possible only on clear moonless nights, so that the duty cycle is equal at best to about 10% of calendar time. Also, with rare exceptions only many particles together can generate enough light to be detected; therefore, it is almost inevitable that extensive air showers should be observed.

Here, again there are three types of experiments.

(i) Investigations of extensive air showers (EAS) of cosmic rays to derive the cosmic ray spectrum and composition.

(ii) Very high-energy (VHE) gamma ray astronomy.

(iii) Observation of Cherenkov light reflected from the Earth's surface by elevated (aircraft-borne) detectors.

---

[1] Invited talk presented at the conference "P.A. Cherenkov and Modern Physics" (Moscow, June 22-25, 2004) commemorating P.A. Cherenkov centenary.

It is very interesting that a single man initiated all three of these lines of research! A.E. Chudakov was the one who started nearly everything in this area. Therefore, it was quite justifiable that Chudakov should be the one to have written Cherenkov's obituary cited in the epigraph.

These three types of experiments differ in the area of light collectors. In the first case PM-tubes without mirrors or with small mirrors are used. The second group of experiments uses large-area mirrors as light collectors (10 m or more in diameter for most modern Cherenkov telescopes). Finally, in the third case the light-reflecting area is represented by a snowy ground surface, and the particular acceptance area depends on the viewing angle and altitude of an instrument.

In what follows, the Chudakov's contributions to the development of the air Cherenkov method will be considered, as well as recent achievements and current state of the art.

## Investigations of extensive air showers

> "…He described his idea in some detail, dating it to 1955-57, the time he made pioneering measurements on atmospheric Cherenkov radiation from EAS."
>
> J. Linsley (Linsley, 2001)

P.M. Blackett was the first who paid attention to the fact that Cherenkov radiation generated by high-energy charged particles could be observed not only in dense media but in air as well (Blackett, 1949). In 1952, Galbraith and Jelley (Galbraith and Jelley, 1953) discovered short light flashes on the background of the nigh-sky glow. These flashes were shown to be associated in some cases with extensive air showers of cosmic rays (Jelley and Galbraith, 1953; Nesterova and Chudakov, 1955) and interpreted as caused by Cherenkov radiation accompanying the extensive air showers.

Galbraith and Jelley used in their experiments one small Cherenkov detector. Chudakov's experiments carried out in the Pamirs Mountains (Chudakov and Nesterova, 1958; Chudakov et al., 1960) included many well separated detectors, some o them with mirrors and some without. Chudakov has realized the idea of calorimetric measurements of the energy of air shower cascades and measured the energy spectrum of primary cosmic rays in a wide range applying the technique of fast oscillography in eight channels simultaneously. It was the first experiment where the EAS Cherenkov radiation was studied in great detail, and it is quite true that in the above quotation of J. Linsley these experiments are referred to as pioneering. (In actual fact, this can be said about almost all works performed by Chudakov throughout his entire life.) The results of these experiments were world-best at least for two decades, and many interesting experimental methods were suggested by him in this Pamirs period, like the use of a small spark as a light source for calibration of photomultipliers. It is also of great (historical) interest that when preparing these experiments in 1953 Chudakov began to study the luminiscence of air and other gases irradiated by relativistic electrons.[2] The aim of these experiments was to check that the ionization glow of air would not be an obstacle for observation of Cherenkov light. He found the ionization glow to be rather weak and negligible for Cherenkov observations. But Chudakov immediately understood that the isotropy of this radiation could be used in experiments of another type in order to observe extensive air showers from a large distance. This idea was realized much later in fluorescence experiments (the famous Fly's Eye detector was the first), and now the detectors of this type are being developed both for ground-based (the Auger project) and for satellite (EUSO) experiments.

---

[2] The experiment was made at various pressures, and, reducing pressure to zero, Chudakov discovered that some signal still existed at zero pressure. Putting additional metal foils into the beam of electrons he proved this signal to be the result of transition radiation predicted by V.L. Ginzburg and I.M. Frank in 1945. This was the first experimental observation of the transition radiation.

After the death of Chudakov John Linsley wrote in 2001 to the author: "I tried … to get clarification from Chudakov himself in his later years about an idea that *apparently came to him before it came to others*: to observe EAS by means of atmospheric scintillation. In a well-known remark of his at the 1962 Interamerican Symposium in La Paz, Bolivia, published in the Proceedings, he described his idea in some detail, dating it to 1955-57, the time he made pioneering measurements on atmospheric Cherenkov radiation from EAS" (Linsley, 2001). Almost a decade later Chudakov published his suggestion in a paper written together with his student (Belyaev and Chudakov, 1966) and completely unknown to the people involved in modern fluorescence experiments.

Modern experiments usually include many detectors of Cherenkov light distributed over a large area. The biggest number of Cherenkov detectors was used up to now by the BLANCA array (144 detectors with an average separation of 35-40 m). Each BLANCA detector contains a large Winston cone which concentrates the light striking an 880 cm$^2$ entrance aperture onto a photomultiplier tube. The concentrator has a nominal half-angle of 12.5° and truncated length of 60 cm. The Winston cones were aligned vertically. The minimum detectable density of a typical BLANCA unit is approximately one blue photon per cm$^2$.

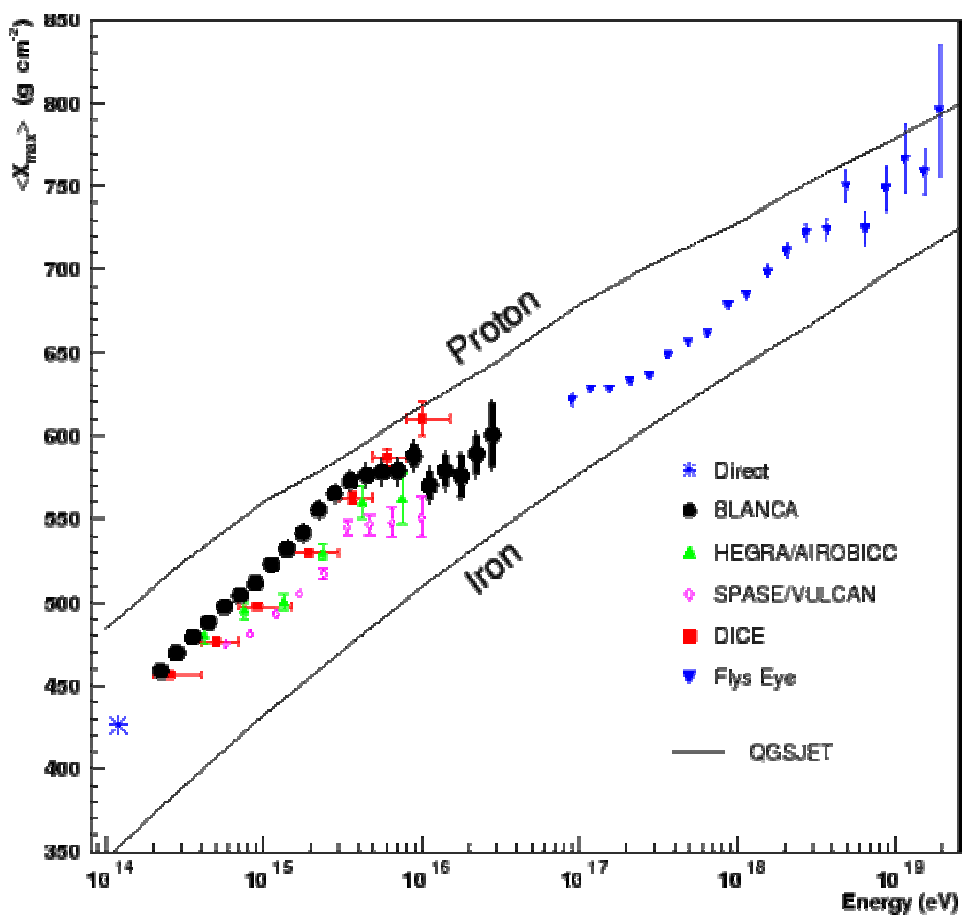

Fig. 1. The data on shower maximum depth as measured in Cherenkov experiments and the Fly's Eye experiment. Cosmic ray composition most probably changes with energy.

This array was located at the same place as the CASA air shower array (a lattice of 900 scintillation detectors with a step of 15 m), and the data of these two experiments were jointly analyzed. The CASA BLANCA (Fowler et al., 2001) experiment is an example of a combination of two independent arrays. Moreover, the third independent Cherenkov experiment DICE (Boothby et al., 1997) was also carried at the same place (DICE is rather unusual experiment of this type since it used only two rather big mirrors). Similar situation takes place at Canari Islands where the HEGRA collaboration constructed first an air shower array with scintillation detectors, then AIROBICC array of Cherenkov detectors (Karle et al., 1995), and Cherenkov telescopes for VHE gamma ray astronomy (Pühlhofer et al., 2003).

An experiment to measure the electron, muon and air-Cherenkov components of 1 PeV air showers has been established at the geographic South Pole (Dickinson et al., 2000). The experiment comprises: the SPASE-2 scintillator array, the VULCAN air Cherenkov array, and the deep under-ice muon detector AMANDA. Simultaneous measurement of the electron size, high energy (>500 GeV) muon content and lateral distribution of Cherenkov light should allow a composition measurement that is relatively insensitive to model assumptions. The SPASE-2 data are used to determine the shower core position and direction. Digitized waveforms from the nine-element VULCAN array are used to reconstruct the lateral distribution of Cherenkov light.

Thus, such coordinated measurements of extensive air showers by particle and light detectors are a general tendency in all modern experiments. Nevertheless, until now there exist 'purely' Cherenkov arrays, for example, Tunka (Gress et al., 1999).

The lateral distribution of Cherenkov light from air showers is closely related to their longitudinal development and can be used to provide an indirect measurement of the depth of shower maximum and primary energy. So, all above-listed experiments carry out the measurements of the energy spectra and the shower maximum position as a function of energy. Figure 1 presents the world data on the shower maximum position including some data obtained with fluorescent detector Fly's Eye at higher energies. The curves for proton and iron primaries are the result of calculation and are model-dependent (particular curves of Fig. 1 are calculated using the QGSJet model). The data of Fig. 1 are not quite consistent, but it seems obvious that the cosmic ray composition changes with energy.

One should also add that Cherenkov detectors are often used in air shower experiments for calibration purposes. For example, Cherenkov detectors in the Yakutsk air shower array do not merely coexist with particle detectors, but Cherenkov detectors are even installed on scintillation detectors. The Yakutsk team uses so-called calorimetric method for estimating the energy of extensive air showers (the energy is determined as a sum of different components of EAS). The Cherenkov light signal represents the energy dissipated above the level of observation and comprises about 75% in the total energy balance. Thus, the data of Yakutsk air shower array depend on the Cherenkov calibration substantially (Glushkov et al., 2003), though independent Cherenkov measurements are also made (Ivanov et al., 2003).

## VHE gamma ray astronomy

The paper (Zatsepin and Chudakov, 1961) in which a new method of searching for high-energy gamma rays of cosmic origin was suggested was published in 1961. And at the moment of publication the first gamma ray telescope was already operating at Catsively, Crimea (Fig. 2). It was designed by Chudakov with a very small group of people. This was not only the beginning of the air Cherenkov method in gamma ray astronomy. This Cherenkov telescope was the first instrument ever specially made for observations in gamma ray astronomy.

Generally, the result of this first experiment was mainly negative: no celestial sources of TeV gamma rays were discovered (Chudakov et al., 1965). First of all, it was unclear what celestial objects should be selected for observations. Radio galaxies were considered to be most promising as potential sources, and they were mainly observed by Chudakov and his team. Due to some bad luck, one of them (Cygnus A) showed statistically significant excess in one of the first runs. So, a lot of time was spent observing this object, in a hope to confirm the signal that finally disappeared.

However, one gamma ray emitter was known at that time, and at least one result of fundamental importance was obtained by Chudakov from the upper limit on the flux of high energy gamma rays derived from observations of this object. This is the famous Crab Nebula source. It was generally believed at that time that the synchrotron radiation in the Crab Nebula was produced by electrons of secondary origin (produced by pions generated in proton-proton collisions via $\pi \to \mu \to e$ decay). If so, one would expect a significant gamma ray flux from decays of neutral pions. The upper limit obtained by Chudakov was a proof of direct acceleration of electrons in the Crab Nebula (and hence in other similar objects).

In any case, this pioneering experiment laid the foundation for the method and its further development. Many years later former Astronomer Royal for Britain Sir Arnold Wolfendale presented to Chudakov his monograph on gamma ray astronomy with the following (fully correct) autograph inscription: "To the founding father". It is a great pity that now gamma ray astronomers are ignorant about this father. There is no name of Chudakov in a huge review that was written by R. Ong (Ong, 1998) and included as many as 413 items in the list of references. More recent and shorter review by E. Lorentz (Lorentz, 2003) also ignores the pioneers (G.T. Zatsepin and A.E. Chudakov) who suggested the very idea of Cherenkov astronomy.

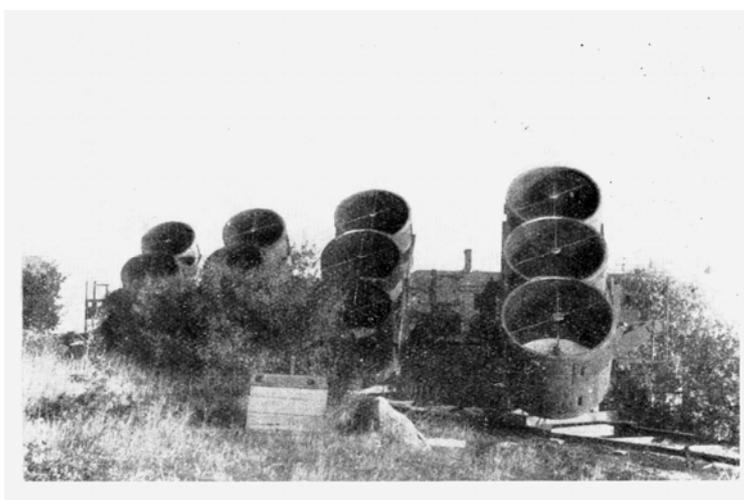

Fig. 2. First gamma ray telescope that was constructed by Chudakov and operated in Catsiveli, Crimea in 1960-1963.

The first gamma ray telescope constructed by Chudakov included 12 mirrors with a diameter of 1.5 m, and it stopped observation in 1963. To illustrate the current state of the art at that time it is sufficient to take into account that the second telescope (started to operate in 1965) had two mirrors with a diameter of 0.9 m (Long et al., 1966). And it was still much later that the first large diameter (10 m) telescope of the Whipple observatory began to be designed. This telescope turned out to be the first successful instrument of gamma ray astronomy: in the middle of 1980s the signal from the Crab Nebula was unequivocally detected with a good statistical accuracy (Weekes et al., 1989). The most important in this success was the so-called imaging technique. A mosaic of PM-tubes was placed in the focal plane of the mirror, and each shower produced a distributed signal in many PM-tubes (Cherenkov image of a shower). The cuts over various parameters of such images allowed one to extract the gamma ray signal from the background of usual extensive air showers generated by cosmic ray nuclei.

The main tendencies in the development of Cherenkov gamma ray telescopes are as follows. (i) The mirror diameters increase, and though there is a natural limitation, the largest mirror used in the MAGIC telescope (Baixeras et al., 2004) has a diameter of 17 m. The natural consequence of increasing mirrors is the reduction of the energy threshold of telescopes (the first Chudakov's telescope has a threshold of $5 \cdot 10^{12}$ eV, while now 100 GeV is more or less standard level and MAGIC team is going to reach even 50 GeV). (ii) The number of PM-tubes in the focal plane also increases and can reach about 500 in modern telescopes (the Whipple telescope started from 19 and then 37 PM-tubes). This also results in an increase of the field of view of the telescopes, and this improves the situation with the background subtraction. (iii) Most modern air Cherenkov telescopes make stereoscopic observations using several large mirrors. The Japanese-Australian telescope CANGAROO-III in Australia (Kabuki et al., 2003) and German telescope HESS in Namibia (Bernlöhr et al., 2003) have 4 mirrors separated by about 100 m. The project VERITAS (Weekes et al., 2002) includes seven 10-12 m reflectors similar to that used in the first Whipple telescope. Among these telescopes of last generation only MAGIC has a single reflector, but recently it has been announced that the twin MAGIC-II telescope is under construction. There are many older and smaller telescopes like French telescope CAT in Pirenei Mountains (Barrau et al., 1998) and German HEGRA telescopes at Canari Islands (Pühlhofer et al., 2003), but four above mentioned telescopes (CANGAROO-III, HESS, VERITAS, and MAGIC) are the most powerful and modern instruments of present-day Cherenkov gamma ray astronomy. They are rather uniformly distributed over the globe (two in the northern and two in the southern hemisphere). It should be pointed out that these systems with multiple light collectors can be easily extended. For example, the HESS collaboration plans a combination of multiple (up to 16) telescopes to a large system to increase the effective detection area for gamma rays. Their proposed system should finally provide for a detection threshold of about 40 GeV, full spectroscopic capability above 100 GeV, an angular resolution for individual showers of 0.1 degrees, and an energy resolution of about 20%. It will allow one to explore gamma-ray sources with intensities at a level of a few thousands of the flux of the Crab Nebula.

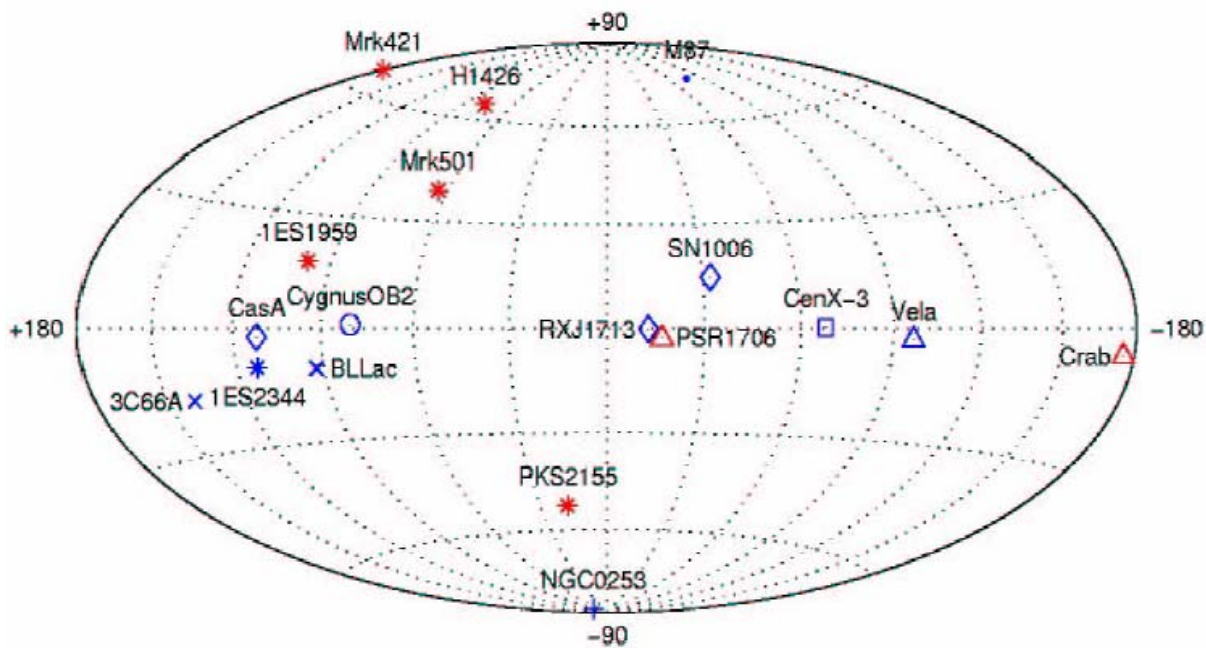

Fig. 3. The map of the sky in TeV gamma rays. All sources are detected by Cherenkov gamma ray telescopes (Weekes, 2003).

A very interesting new development in air Cherenkov gamma ray astronomy is the use of solar power plants as Cherenkov telescopes. A great number of computer-controlled solar concentrators are used at such plants only during the day. In the night time one can use these light reflectors (a part or all of them) in order to track a potential celestial source and to concentrate the Cherenkov light on a light receiver (most frequently, a secondary mirror) placed at the solar tower. The multi-pixel PM-tube camera is usually installed at the focal plane of the secondary mirror. At the moment, there are several successfully operating experiments of this type: American experiments STACEE in New-Mexico (Ong et al., 1996, Hanna et al., 2002) and Solar-II in Barstow, California (Zweerink et al., 1999), French experiment CELESTE (Giebels et al., 1998), and Spanish experiment GRAAL (Arqueros et al., 2002).

Until now the VHE range was separated from the satellite high-energy gamma ray astronomy domain by a certain gap (the upper limit of energy for the EGRET instrument onboard the CGRO was about 30 GeV). The MAGIC energy range is promised to be rather near to this value and the new project CHESS for the optical-infrared Japanese telescope SUBARU (Asakhara et al., 2002) with a 8.3 m mirror plans to reduce the energy threshold of ground-based measurements down to 10 GeV. Thus, the gap between satellite and ground-based observations will be closed.

What about the results obtained at present in the VHE gamma ray astronomy? There are more than a dozen of sources reliably detected. The total number of sources is not very strictly determined quantity, because some sources are marginally detected, many telescopes observe various sources at different places with rather short duty cycle (and some of the sources are strongly variable). Nevertheless, a catalog of these sources (Weekes, 2003) includes about 10 galactic and 8 extragalactic sources. A half of these 18 sources were detected by at least two groups at a $5\sigma$ level). All sources are identified (as opposed to the EGRET catalog of gamma ray sources at lower energy) and represent rather diverse classes of objects (AGN, supernova remnants, radio galaxy, starburst galaxy, binary source, and OB association). The map of TeV gamma ray sky is shown in Fig. 3. Detailed discussion of these results is beyond the scope of this paper. I would like only to emphasize that the study of cosmic accelerators of particles is of utmost importance for astrophysics, and Cherenkov VHE gamma ray astronomy is a well established branch of astronomy with very good prospects for the future. The number of newly discovered sources increases every year.

## Observation of reflected Cherenkov light

This experimental method was suggested by Chudakov in 1972 (Chudakov, 1972). The idea was to measure the energy spectrum of giant extensive air showers (for a particular version of the experiment Chudakov estimated that at energy $10^{18}$ eV one can reach the signal to noise ratio of about 5) by a small light detector installed onboard an airplane flying during the polar night over northern territories of Russia (tundra). One advantage of this experiment would be continuous observation for a long time (which is unusual for air Cherenkov observations). This suggestion, however, has not been realized by Chudakov himself who was involved in many other experiments. The attempts to implement his idea started from an experiment (Castagnoli et al., 1983), in which using a mountain relief of Alps the Cherenkov detectors observed the surface of a glacier from a high point. Another similar experiment was carried out in Tien Shan mountain region (Antonov et al., 1997).

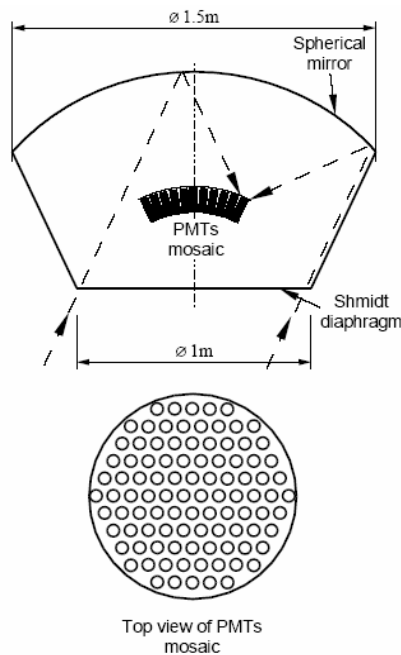

Fig. 4. The layout of the SPHERE detector.

The next step in realization of this idea was made (Antonov et al., 1999) with the same detector placed to a fixed (tethered) balloon (experiment SPHERE-1). This detector represents a spherical concave mirror with a convex matrix of PM-tubes (Fig. 4). At the moment, SPHERE-1 detector measured the energy spectrum of extensive air showers in the interval $10^{16} - 10^{17}$ eV. The same type of detector is planned to be used in a high-altitude balloon in Antarctica (Antonov et al., 2001). This project (experiment SPHERE-2) should have much higher energy threshold (near to $10^{20}$ eV). Also, if realized in a circumpolar region it will allow one to make continuous observations as was originally suggested by Chudakov. These two experiments with balloons at altitudes of 1-3 km (SPHERE-1) and more than 30 km (SPHERE-2) are currently the only real attempts to realize the Chudakov's idea, though the EUSO experiment (Catalano et al., 2003), now in preparation to be installed onboard the International Space Station, also discusses possibilities to detect the reflected Cherenkov radiation in addition to the main purpose of detecting the fluorescent light of EAS.

## Other experiments and ideas

Three lines of research reviewed above are the main tendencies in air Cherenkov method applications. There exist, however, some experiments that stand apart from this classification. First of all, one should point to the experiments where the Cherenkov radiation of not cascades but single particles is detected. This is possible to do only for nuclei with a fairly high charge (for example, iron nuclei). First attempt of such an experiment was made by Sood (Sood, 1983), and in more recent balloon experiment BACH (Seckel et al., 1999) the flux of iron nuclei in primary cosmic rays was measured at the energy of about $10^{13}$ eV. The general appearance of the BACH instrument is shown in Fig. 5.

The idea to combine a measurement of the Cerenkov light produced by the incoming cosmic-ray nucleus in the upper atmosphere with an estimate of the total nucleus energy produced by the extensive air shower initiated when the particle interacts deeper in the atmosphere was discussed by Kieda, Swordy, and Wakely (Kieda et al., 2001). They hope that the emission regions prior to and after the first hadronic interaction can be separated by an imaging Cerenkov system with sufficient angular and temporal resolution. Their Monte Carlo simulations showed that one can reach an expected charge resolution of $\Delta Z/Z < 5\%$ for incident iron nuclei in the region of the "knee" of the cosmic-ray energy spectrum. This technique also has the intriguing possibility to unambiguously discover nuclei heavier than iron at energies above $10^{14}$ eV.

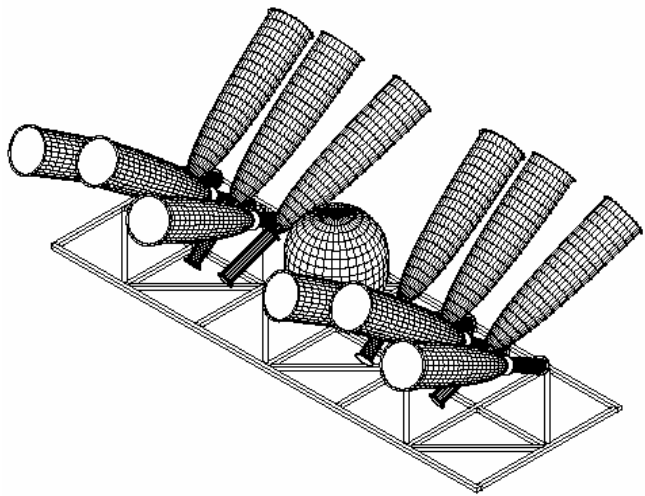

Fig. 5. The instrument of the BACH balloon experiment. The Cherenkov light generated by high-energy primary iron nuclei is collected by light pipes.

A new idea is to construct an air Cherenkov detector for horizontal air showers produced by Earth-skimming and mountain-traversing neutrinos (experiment NuTel). A possibility to detect horizontal air showers generated by $\tau$-neutrinos was widely discussed for new giant air shower arrays. The NuTel collaboration (Hou and Huang, 2001) is the only one that plans to use the Cherenkov detector for this experiment. The possibility to use the fluorescence detector in combination with the Cherenkov detector is also discussed. The collaboration has already found a particular place for this experiment at Hawaii and the work is in progress. Thus, neutrino physics and astronomy is the new application of the air Cherenkov method.

## Conclusion

The radiation first discovered by P.A. Cherenkov is used for a variety of methods in cosmic ray studies. Among them, air Cherenkov methods form a separate area with several lines of research. Nearly all of the latter were first developed or suggested by A.E. Chudakov whose ideas and experimental skill laid the foundation for present-day progress.

The studies of Cherenkov radiation of extensive air showers were from the very beginning closely connected with the studies of the ionization glow of the showers. The experiments detecting one type of emission considered another one as a background. Now there are plans of complementary measurements of both types of emissions.